\documentclass{article}
\usepackage[utf8]{inputenc}
\usepackage{fullpage}
\usepackage{booktabs}
\usepackage{makecell}

\usepackage[english]{babel}
\usepackage{csquotes}
\usepackage{hyperref}

\usepackage{amssymb}
\usepackage{dcolumn}
\usepackage{enumerate}
\usepackage{mathdots}
\usepackage{tabularx}
\usepackage{subfigure}
\usepackage{amsmath}
\usepackage{amsthm}
\usepackage{tikz} 
\usepackage{float}
\usepackage{mathtools}
\usepackage{graphicx}
\usepackage{pgfplots}
\usepackage{pgf,tikz}
\usepackage{mathrsfs}
\usepackage{authblk}
\pgfplotsset{compat=1.17} 

\usepackage{algorithm}
\usepackage{algorithmic}

\usepackage{longtable}
\usepackage{siunitx}

\title{High-Accuracy Machine Learning Techniques for Functional Connectome Fingerprinting and Cognitive State Decoding}

\author[1]{Andrew Hannum}
\author[1]{Mario A. Lopez}
\author[2]{Saúl A. Blanco}
\author[2]{Richard F. Betzel}
\affil[1]{Department of Computer Science, University of Denver}
\affil[2]{Department of Computer Science, Indiana University, Bloomington}
\affil[3]{Department of Psychological and Brain Sciences, Indiana University, Bloomington}

\begin{document}

\maketitle

\begin{abstract}
The human brain is a complex network comprised of functionally and anatomically interconnected brain regions. A growing number of studies have suggested that empirical estimates of brain networks may be useful for discovery of biomarkers of disease and cognitive state. A prerequisite for realizing this aim, however, is that brain networks also serve as reliable markers of an individual. Here, using Human Connectome Project data, we build upon recent studies examining brain-based fingerprints of individual subjects and cognitive states based on cognitively-demanding tasks that assess, for example, working memory, theory of mind, and motor function. Our approach achieves accuracy of up to 99\% for both identification of the subject of an fMRI scan, and for classification of the cognitive state of a previously-unseen subject in a scan. More broadly, we explore the accuracy and reliability of five different machine learning techniques on subject fingerprinting and cognitive state decoding objectives, using functional connectivity data from fMRI scans of a high number of subjects (865) across a number of cognitive states (8). These results represent an advance on existing techniques for functional connectivity-based brain fingerprinting and state decoding. Additionally, 16 different pre-processing pipelines are compared in order to characterize the effects of different aspects of the production of functional connectomes (FCs) on the accuracy of subject and task classification, and to identify possible confounds.
\end{abstract}


\section{Introduction}

Functional connectivity is a measure of the statistical dependency between recorded activity from different sites in the brain~\cite{friston1994functional}. In practice it is operationalized as a correlation measure and usually estimated during the so-called ``resting state", i.e. in the absence of explicit task instruction~\cite{smith2011network}. The complete set of functional connections between all neural elements defines a functional brain network, which can be modeled as a fully weighted and signed graph of nodes (brain regions) and edges (functional connection weights) \cite{bassett2017network,park2013structural}. A large body of work has been compiled, demonstrating that variation in connection weights can be used to accurately predict an individual's cognitive \cite{shirer2012decoding}, clinical \cite{fornito2015connectomics}, and developmental state \cite{gu2015emergence}.

Recently, there has been intense interest in network-based ``fingerprinting" (for example, see~\cite{amico_quest_2018,chiem_improving_2022,FBS22,jalbrzikowski_functional_2020,stampacchia_fingerprinting_2022,sorrentino_clinical_2021}.)
That is, using statistical analysis to discover features of functional brain networks that reliably distinguish one individual from another. Previous research has shown that the functional connectome (FC) can provide a fingerprint of an individual consistent across fMRI scans, and also that they encode aspects of generalized cognitive states in an identifiable way.

There remain, however, key unanswered methodological questions. For instance most studies have not directly assessed the relative performance of different algorithms for fingerprinting, nor have they exhaustively explored the impact of the FC pre-processing pipeline on fingerprinting.

Here, we explore the accuracy and reliability of five different machine learning techniques on subject fingerprinting and cognitive state decoding objectives using FCs from a Human Connectome Project (HCP) dataset (see~\cite{van2013wu}) with 865 subjects and engaged in 8 different in-scanner tasks. Additionally, four pre-processing pipeline variables producing a total of 16 different pre-processing pipelines are compared in order to characterize the effects of different aspects of the production of FCs on the accuracy of subject and task classification, and in order to identify possible confounds.


The results demonstrate a near-perfect ability of some machine learning techniques to perform both subject fingerprinting and task decoding from FCs in the HCP dataset, providing support for the idea that an FC reveals important characteristics on an individual and their cognitive state, and giving a benchmark against which future techniques can be measured. An additional benefit of these high-accuracy classifiers is the ability to identify which features of the FC, and therefore which connections between which regions of the brain, are regions-of-interest for further investigation of cognitive states. 


\subsection*{Previous work}

Goni and Amico have done work on fingerprinting/identification of individuals based on their connectivity\cite{abbas_geodesic_2021, amico_quest_2018}. Their work builds on the methods of Finn et al.\cite{finn_functional_2015} which are based on the use of a Pearson correlation coefficient (or some other distance function) to identify the "closest match" of a target scan. Previous results have achieved subject identification accuracy of $>90\%$ in limited cases such as matching between scans engaged in the same task (\emph{rest}$\rightarrow$\emph{rest}).

For cognitive state decoding, Shirer et al.~\cite{shirer_decoding_2012} showed a classifier based on a 90-region FC that was able to distinguish between 4 cognitive states with 85\% accuracy on a cohort of previously-unseen subjects. More recently, Wang et al.~\cite{wang_decoding_2019} showed that a deep neural-network (DNN) architecture can achieve 93.7\% accuracy when distinguishing between 7 cognitive states using raw fMRI data as inputs. Huang et al.~\cite{huang_design_2021} achieved similar accuracy with another deep learning model. Saeidi et al.~\cite{saeidi_decoding_2022} proposed a graph neural-network (GNN) model that achieves accuracy of 97.7\% when distinguishing between 7 cognitive states.

In other related work, some authors have looked at cognitive state decoding with goals different from our own~\cite{gao_decoding_states,gupta_decoding_2022}.

\subsection*{Our contribution}

Our main contribution establishes that Linear Discriminant Analysis classifier can significantly outperform other common machine learning approaches on a subject fingerprinting objective, achieving identification accuracies of $\sim99\%$ for a set of 865 subjects with target scans of cognitive states unseen in classifier training.

In addition, we present a detailed description of the effects of four pre-processing steps, and we identify small sub-networks of brain regions (via feature selection of connectome edges, and LDA coefficients) that underpin the high accuracies achieved for both subject fingerprinting and task decoding - all of which provide direction for future investigation.

\section{Methods}

\subsection{Data description and classification objectives} 

\textbf{Dataset.} The original dataset consists of fMRI scans from 865 subjects released through the Human Connectome Project. For each subject there are 2 scans for each of 7 active tasks (\emph{emotion}, \emph{gambling}, \emph{language}, \emph{motor}, \emph{relational}, \emph{social}, \emph{working memory}) and 4 scans taken during a resting state task (\emph{rest}), for a total of 18 scans per subject. In order to remove any possible effects of an unbalanced dataset due to the surplus of rest scans, only 2 rest scans per subject were used during the experiments, 
for a total of 16 scans per subject. No personal identifiable information is contained in the dataset.

A functional connectivity matrix is produced from an individual fMRI scan via a pipeline that includes brain-region parcellation, confound regression, and calculation of a correlation coefficient matrix from some subset of the fMRI time-series frames.


\textbf{HCP Functional Pre-processing} (see ~\cite{Sporns21}). The images in the HCP dataset were minimally preprocessed as described in~\cite{Glasser13}. Briefly, each image was corrected for gradient distortion and motion, and aligned to a corresponding T1-weighted (T1w) image with one spline interpolation step. This was further corrected for intensity bias and normalized to a mean of 10,000 and projected to a 32k\_fs\_LR mesh, excluding outliers, and aligned to a common space using a multi-modal surface registration~\cite{robinson2014msm}.

\textbf{Parcellation Pre-processing} (see ~\cite{Sporns21}). A 
parcellation consisting of 200 areas on the cerebral cortex was designed so as to
optimize both local gradient and global similarity measures of the fMRI signal~\cite{schaefer2018local} (Schaefer200). The parcellation nodes were mapped to the Yeo canonical functional networks~\cite{Yeo11}. For the HCP dataset, the Schaefer200 is openly available as a CIFTI file in 32k\_fs\_LR space. These tools utilize the surface registrations computed in the recon-all pipeline to transfer a group average atlas to subject space based on individual surface curvature and sulcal patterns. This method rendered a T1w space volume for each subject. 
The parcellation was resampled to 2 mm T1w space for use with functional data. The same process can be used for other resolutions (e.g. Schaefer100).

\textbf{Functional Network Pre-processing} (see ~\cite{Sporns21}). All the BOLD images were linearly detrended, band-pass filtered (0.008 - 0.08 Hz)~\cite{parkes2018evaluation}, confound regressed and standardized using Nilearn’s signal.clean, which removes confounds orthogonally to the temporal filters~\cite{lindquist2019modular}. The confound regression employed~\cite{Satterthwaite12} has been shown to be a relatively effective option for reducing motion-related artifacts~\cite{parkes2018evaluation}. Following pre-processing and nuisance regression, residual mean BOLD time series at each node were recovered.

\textbf{Classification Objectives.} In each experiment, a machine learning classifier was trained to accomplish one of two objectives: subject identification (fingerprinting) or task identification (decoding).

\textbf{(i) Subject Fingerprinting.} For 
this objective, the classifier was trained on scans from all subjects simultaneously, across 7 tasks for each subject - a total of 14 scans per subject. For each subject, one of the 8 tasks was randomly selected, and the 2 scans for that task were withheld to be used as the verification dataset. This ensures that scans from all subjects have been seen by the classifier during training, but that none of the scans in the verification set match both the subject and task of any scan seen during training (i.e. during validation, the classifier will necessarily be matching the subject, and not the task). The accuracy of the subject fingerprinting classifier is a measure of the percentage of the verification scans whose subject was correctly predicted using the trained classifier.

\textbf{(ii) Task Decoding.} For the task decoding objective, the classifier was trained on scans from a subset of subjects (see section \ref{training_validation}), across 8 tasks - a total of 16 scans per subject. The verification dataset consisted of the scans from the remaining subset of subjects across all tasks. This ensures that scans in the verification set are from tasks that the classifier has seen with equal distribution, but from subjects that the classifier has never seen before (i.e. during validation, the classifier will necessarily be matching the task, and not the subject). The accuracy of the task decoding classifier is a measure of the percentage of the verification scans whose task was correctly predicted using the trained classifier.

\subsection{Classifier Architectures}

Five classifier architectures were tested across the two objectives:
	
\begin{enumerate}
\item A linear discriminant analysis (LDA) classifier, whose results give, for each class, a set of linear coefficients to the input vector that can be used to draw conclusions (and produce visualizations) about the relative importance of components of the input towards belonging to a particular class.
\item A multi-layer perceptron neural network (NN) classifier, whose training can model higher-order functions.
\item A support-vector machine (SVM) classifier, a common machine learning approach to classification with high-dimensional data. Many recent results in human brain functional connectomics focus on SVM as their predictive tool \cite{he_early_2018,mao_resting-state_2022,watanabe_disease_2014,yang_whole-brain_2017}.
\item A nearest-centroid (NC) classifier, whose results can be used to draw simple inferences about the high-dimensional spatial relationship between the classes.
\item A correlation-based (CORR) classifier, the classical method for subject fingerprinting on functional connectomes known to achieve high accuracy on similar datasets \cite{finn_functional_2015}.
\end{enumerate}

For further experiments, including comparison of pre-processing pipelines, feature selection, and analysis of high-uniformity subjects, we focused on either the highest-accuracy classifier (LDA), or the two classifiers that maintained the highest average accuracy across both objectives (LDA and NN).

\subsection{Pre-processing Pipelines}

Four pre-processing pipeline variables were tested in order to determine the robustness of the results to differences in data preparation, including:

\begin{itemize}
\item With or without global signal regression \cite{murphy_Towards_a_consensus_GSR}: removal of the average signal intensity of the fMRI time series through linear regression.
\item With or without task regression \cite{cole_task_2019}: removal of the first-order effect of task-evoked activations.
\item Schaefer-Yeo 100-region vs 200-region parcellation \cite{schaefer2018local}: the number of regions in the brain atlas used to produce the FCs.
\item Truncation of the \emph{rest} scan time-series frames to be consistent with the shorter length of \emph{task} scans. The scan lengths are truncated for consistency across all cognitive states.
\end{itemize}



\section{Results}


In this section we present and discuss: 

\begin{itemize}
    \item The accuracy of the five machine learning architectures on subject fingerprinting and task decoding.
    \item The effects of training/validation split on classifier accuracies.
    \item The effects of the four pre-processing pipelines on classifier accuracies.
    \item Three methods of feature selection for identifying brain region connections that are useful for fingerprinting and decoding.
    \item Analysis of the high-dimensional clustering of cognitive states.
\end{itemize} All the experiments were run using an AMD Ryzen 7 3700X 3.6 GHz 8-Core CPU.

\subsection{Classifier Architecture Accuracies}
\begin{figure*}[t]
\centering
\footnotesize
\noindent\begin{tabular}{ cccccccccccccc  }
\toprule
 \multicolumn{4}{c}{Pipeline} & \multicolumn{5}{c}{\thead{Fingerprinting\\ Accuracy}} &
 \multicolumn{5}{c}{\thead{Task Decoding\\ Accuracy}}\\
 \cmidrule(lr){1-4}
 \cmidrule(lr){5-9}
 \cmidrule(lr){10-14}
 \thead{ Regions} & \thead{ Global\\ Reg.} & \thead{ Task\\ Reg.} & \thead{\ Trunc.} & \thead{ LDA} & \thead{ NN} & \thead{SVM} & \thead{NC} & \thead{CORR} & \thead{ LDA} & \thead{ NN} & \thead{SVM} & \thead{NC} & \thead{CORR} \\
 \cmidrule(lr){1-4}
 \cmidrule(lr){5-9}
 \cmidrule(lr){10-14}
100 & Yes & No   & No    & 0.979 & 0.645 & 0.309 & 0.232 & 0.061 & 0.995 & 0.992 & 0.995 & 0.895 & 0.742 \\
100 & Yes & Yes  & No    & 0.938 & 0.573 & 0.312 & 0.297 & 0.069 & 0.899 & 0.944 & 0.975 & 0.768 & 0.415 \\
100 & No  & No   & No    & 0.974 & 0.458 & 0.176 & 0.151 & 0.059 & 0.988 & 0.994 & 0.991 & 0.691 & 0.756 \\
100 & No  & Yes  & No    & 0.850 & 0.304 & 0.171 & 0.141 & 0.067 & 0.904 & 0.943 & 0.957 & 0.509 & 0.526 \\
 \cmidrule(lr){1-4}
 \cmidrule(lr){5-9}
 \cmidrule(lr){10-14}
200 & Yes & No   & No    & 0.994 & 0.839 & 0.609 & 0.477 & 0.153 & 0.979 & 0.996 & 0.998 & 0.914 & 0.701 \\
200 & Yes & Yes  & No    & 0.983 & 0.765 & 0.565 & 0.588 & 0.190 & 0.882 & 0.953 & 0.986 & 0.793 & 0.345 \\
200 & No  & No   & No    & 0.995 & 0.801 & 0.420 & 0.297 & 0.141 & 0.977 & 0.994 & 0.993 & 0.696 & 0.714 \\
200 & No  & Yes  & No    & 0.947 & 0.703 & 0.353 & 0.275 & 0.131 & 0.876 & 0.953 & 0.976 & 0.544 & 0.482 \\
 \cmidrule(lr){1-4}
 \cmidrule(lr){5-9}
 \cmidrule(lr){10-14}
100 & Yes & No   & Yes    & 0.914 & 0.591 & 0.309 & 0.204 & 0.071 & 0.989 & 0.991 & 0.995 & 0.881 & 0.742 \\
100 & Yes & Yes  & Yes    & 0.779 & 0.529 & 0.329 & 0.284 & 0.076 & 0.899 & 0.946 & 0.975 & 0.751 & 0.415 \\
100 & No  & No   & Yes    & 0.873 & 0.428 & 0.214 & 0.132 & 0.060 & 0.986 & 0.990 & 0.991 & 0.670 & 0.756 \\
100 & No  & Yes  & Yes    & 0.601 & 0.301 & 0.136 & 0.142 & 0.064 & 0.901 & 0.941 & 0.957 & 0.504 & 0.526 \\
 \cmidrule(lr){1-4}
 \cmidrule(lr){5-9}
 \cmidrule(lr){10-14}
200 & Yes & No   & Yes   & 0.997 & 0.571 & 0.613 & 0.446 & 0.147 & 0.983 & 0.996 & 0.998 & 0.903 & 0.701 \\
200 & Yes & Yes  & Yes   & 0.976 & 0.584 & 0.582 & 0.551 & 0.202 & 0.884 & 0.953 & 0.986 & 0.781 & 0.345 \\
200 & No  & No   & Yes   & 0.990 & 0.498 & 0.406 & 0.291 & 0.144 & 0.976 & 0.994 & 0.993 & 0.679 & 0.714 \\
200 & No  & Yes  & Yes   & 0.929 & 0.417 & 0.352 & 0.275 & 0.136 & 0.877 & 0.948 & 0.976 & 0.527 & 0.482 \\
\bottomrule
\end{tabular}
\caption{Validation accuracy scores of all classifier architectures and all pipelines.}\label{tables_all}
\end{figure*}

\begin{figure*}[t]
\centering
\includegraphics[width=0.45\textwidth]{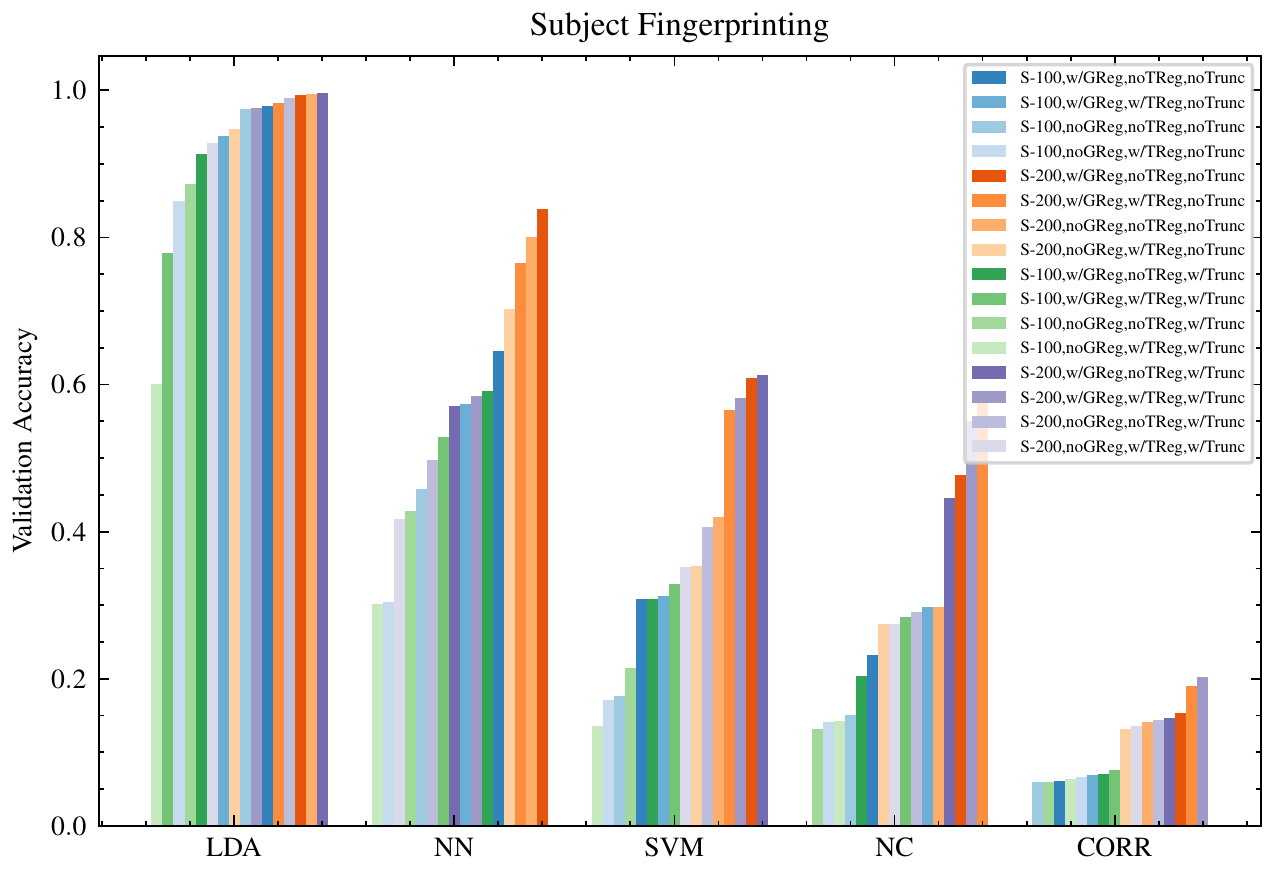}
\includegraphics[width=0.45\textwidth]{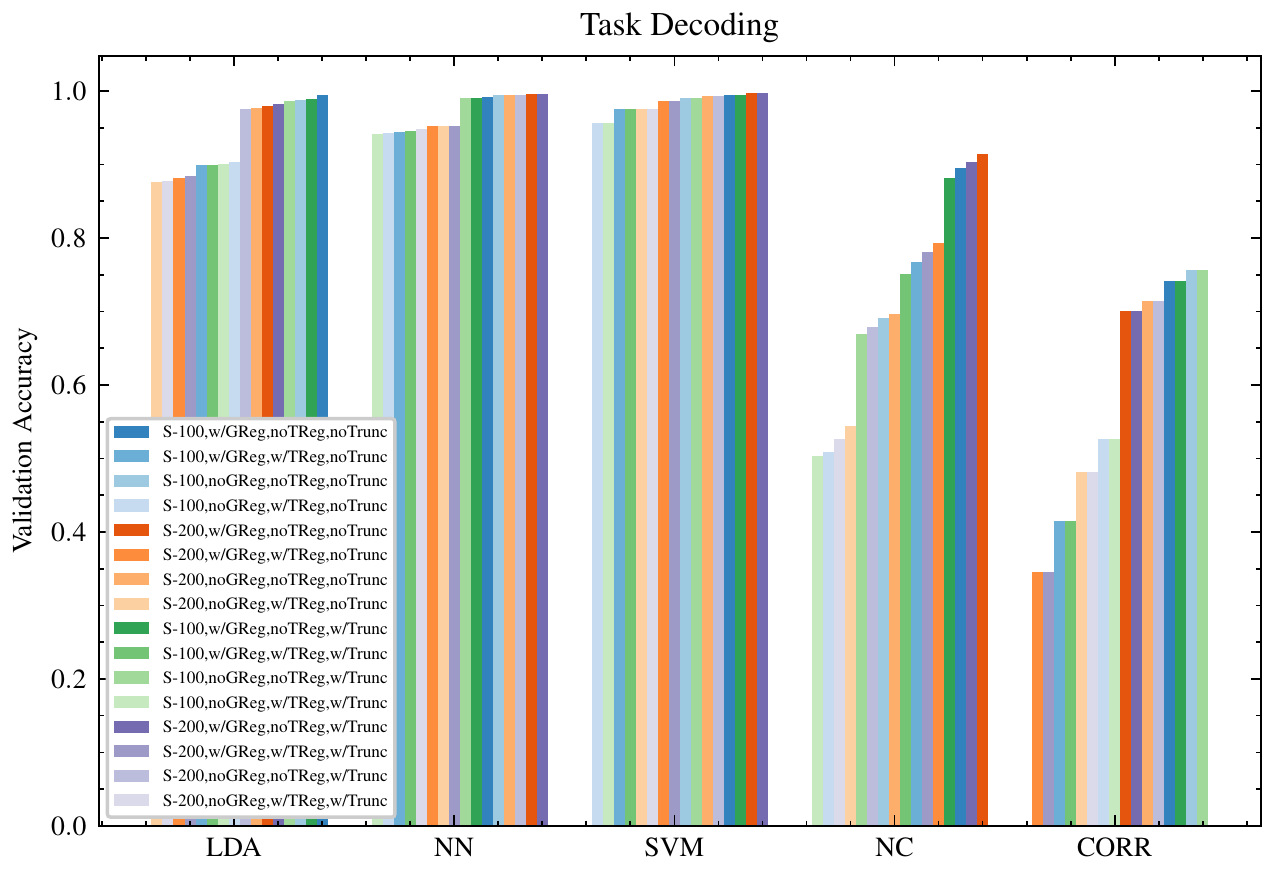}
\caption{Comparison of validation accuracy scores for subject fingerprinting classifiers and task decoding classifiers.}\label{classifiers_all_visual}
\end{figure*}

\subsubsection{Linear Discriminant Analysis Classifier}

The LDA classifier achieved high accuracy scores across both objectives, and across most pre-processing pipelines (see Figure \ref{tables_all}). Overall it appears to be more sensitive to the effects of variations in the pre-processing pipeline than the NN classifier.

The coefficients obtained from the LDA classifier (Fig. \ref{heatmaps}) provide some validation for the means of the class differentiation by giving a relative weighting of input feature importance to classification. Mapping the input features to known brain system labels can be used to verify that the classifier is ``paying attention" to features from areas of the brain known to be associated with the task from a particular class. For example, the connections between areas of the motor cortex are the most important features as determined by the LDA coefficients when predicting belonging to the \emph{motor} or \emph{rest} task, which matches expectations from systems-based perspective of the brain regions. 

\subsubsection{Feed-Forward Neural Network Classifier}

Figure \ref{tables_all} also shows that the NN classifier achieved higher and more consistent accuracy than the LDA classifier on the task decoding objective. However, the NN classifier performed markedly worse on the subject fingerprinting objective.

We tested a limited selection of NN hyperparameters: a three-layer network with a single hidden layer of sizes between 100 and 800, rectified linear activation functions, a constant learning rate of 0.001, and the ``adam'' stochastic gradient-based optimizer~\cite{KJ15}. Our limited experiments showed no improvement of accuracy with larger hidden layer sizes, nor an increased number of hidden layers. Nevertheless, it's possible that the worse performance of the NN compared to LDA on subject fingerprinting could be improved with better selection of NN hyperparameters (larger or more hidden layers, different learning rates, different activation functions, etc.), by another choice of NN architecture, or by training with larger datasets.

As a point of comparison, an existing approach\cite{wang_decoding_2019} using a deep neural-network (DNN) architecture was able to achieve up to $\sim$94\% accuracy on a similar task decoding objective from among 7 tasks. Their DNN was based on convolution using the raw fMRI data as inputs. That our simple three-layer NN is able to achieve better accuracy with the significantly dimension-reduced input of FCs suggests that such a deep neural network may not be necessary for robust results. Additionally, whereas the results of Wang et al. found that certain tasks were easier to decode than others using the DNN, our NN classifier saw essentially no difference in classification accuracy across different tasks.

\subsubsection{Support-Vector Machine Classifier}

The SVM classifier achieved, by a small but noticeable margin, the highest and most consistent accuracy scores of all architectures on the task decoding objective (see Fig. \ref{classifiers_all_visual}), with results that appear more robust to variations in the pre-processing pipeline than the LDA classifier and the NN classier.

On the subject fingerprinting objective the accuracy of the SVM classifier was noticeably worse than both the LDA and NN classifiers, and showed significant sensitivity to the pre-processing pipelines.

\subsubsection{Nearest-Centroid Classifier}

The NC classifier resulted in poor performance across both objectives and all pre-processing pipelines, however it still achieved a level of accuracy high enough to draw some conclusions about the nature of the dataset. For task decoding, accuracy scores of up to $\sim$90\% were seen on some pre-processing pipelines. Given that the measure of the similarity of two connectomes in the NC classifier is the $\ell^2$-norm, the reasonably high accuracy of the results suggests that the dataset exhibits clustering of the classes in high-dimensional space. Results of the NC classifier on subject fingerprinting are similarly poor relative to other architectures, however an ability to correctly identify the subject of a scan approximately half the time, out of 865 subjects, and using only the $\ell^2$-norm, again suggests a relatively strong spatial clustering of scans associated with individual subjects.

\subsubsection{Correlation Classifier}

The Pearson correlation classifier was tested as a point of comparison to known baselines for subject fingerprinting accuracy. Finn et al.\cite{finn_functional_2015} reported achieving high subject identification accuracies by simple calculation of Pearson correlation coefficient between a target scan and a database of test matrices. The predicted identity of the target scan was the one with the highest correlation coefficient (\emph{argmax}).

Their experiments used a 268-node Shen parcellation on a set of 126 subjects from the HCP dataset, with a subject identification objective that was limited to a test database of scans within a single a task. The highest accuracies ($\sim$93\%) were achieved only when the target matrix and test matrix databases were both from scans engaged in the \emph{rest} task (i.e.: \emph{rest}$\rightarrow$\emph{rest}), whereas accuracy suffered for other task pairings (for example \emph{rest}$\rightarrow$\emph{other}). Additionally, they note that the accuracy of identification is sensitive to the choice of parcellation. For example, when using the 68-node FreeSurfer-Yeo atlas, they achieve $\sim$89\% accuracy on \emph{rest}$\rightarrow$\emph{rest} subject identification.

In an attempt to reproduce their results, using 126 random subjects from our dataset, the best accuracy we achieved for the correlation-based identification of \emph{rest}$\rightarrow$\emph{rest} scans was $\sim$81\% using the 200-node Schaefer-Yeo parcellation. It's not clear why they achieved significantly higher accuracy with a similar classification architecture (possible factors may include the parcellation atlas used, pre-processing pipelines, or a difference in the 126 subjects).

Finally, to fully compare the results of the CORR classifier to our best-performing subject fingerprinting classifier (LDA) we extrapolated the technique to our full dataset and task-agnostic identification. Using 779 subjects and a \emph{rest}$\rightarrow$\emph{rest} task objective, the accuracy decreased to $\sim$73\%. Using 779 subjects and an \emph{any}$\rightarrow$\emph{all} task objective (the objective we focus on in our paper for which the LDA classifier achieves $\sim$99\% accuracy), the highest accuracy achieved decreased to $\sim$20\% (Fig. \ref{tables_all}).

\subsection{LDA Feature Importance For Cognitive State Decoding}
\begin{figure*}
\centering
\includegraphics[scale=1]{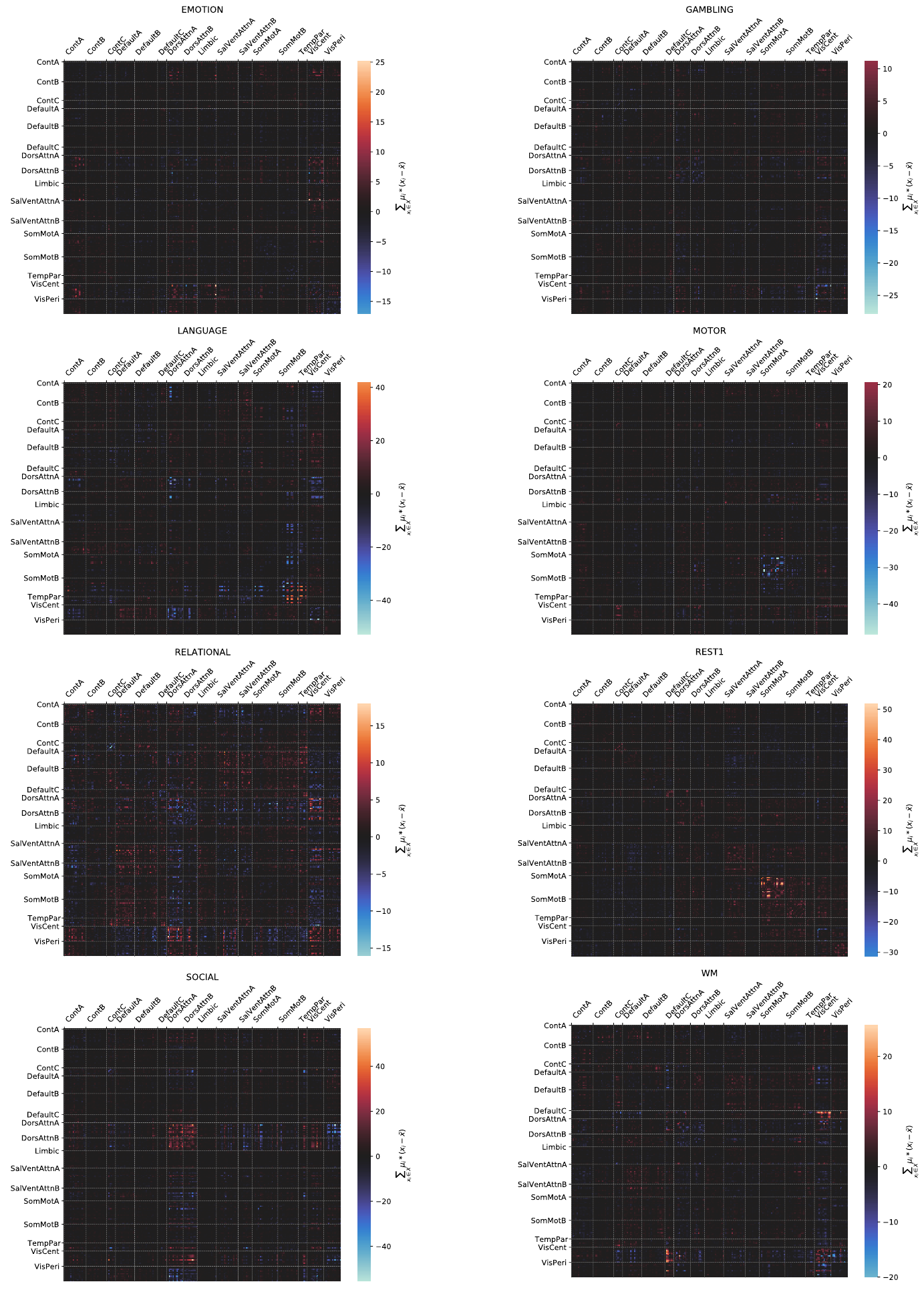}
\caption{Example heatmap visualizations showing the correlation between individual connectome features and particular cognitive states as a function of the coefficients obtained from the LDA classifier. Feature weights here are sum over samples from the class of the feature coefficient times deviation from the mean: $ w_{i} = \sum_{x_{i} \in X} \mu_{i} * (x_{i} - \bar{x}) $. Connectome features are sorted by the associated brain systems.}\label{heatmaps}
\end{figure*}

We produced visualizations (see Figure \ref{heatmaps}) showing the correlation between individual connectome features and particular cognitive states as a function of the coefficients obtained from the LDA classifier.

\subsection{Effects of Training/Validation Split on LDA Classifier}\label{training_validation}

\begin{figure}[H]
\centering
\includegraphics[scale=0.6]{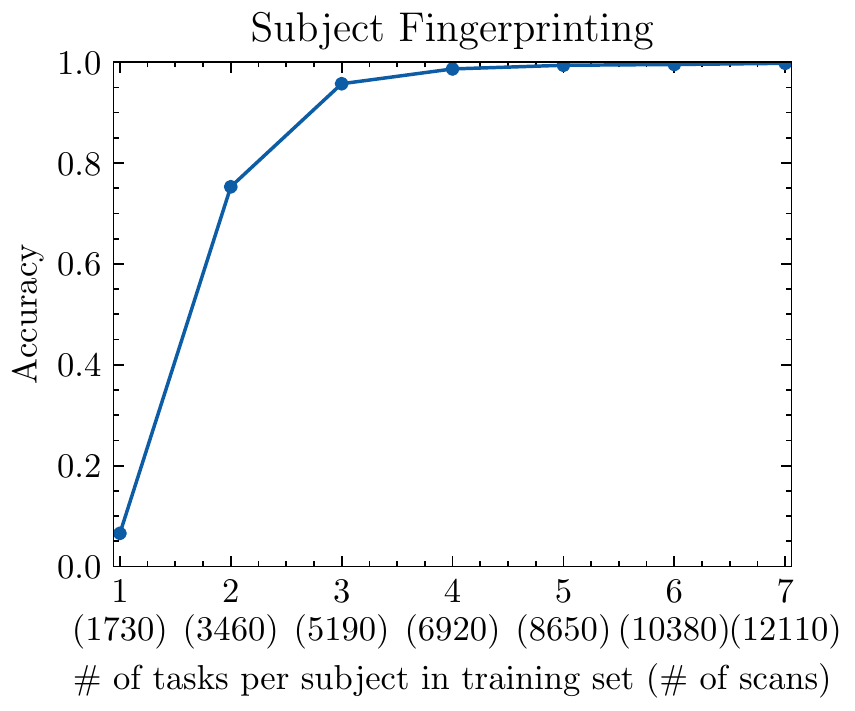}
\includegraphics[scale=0.6]{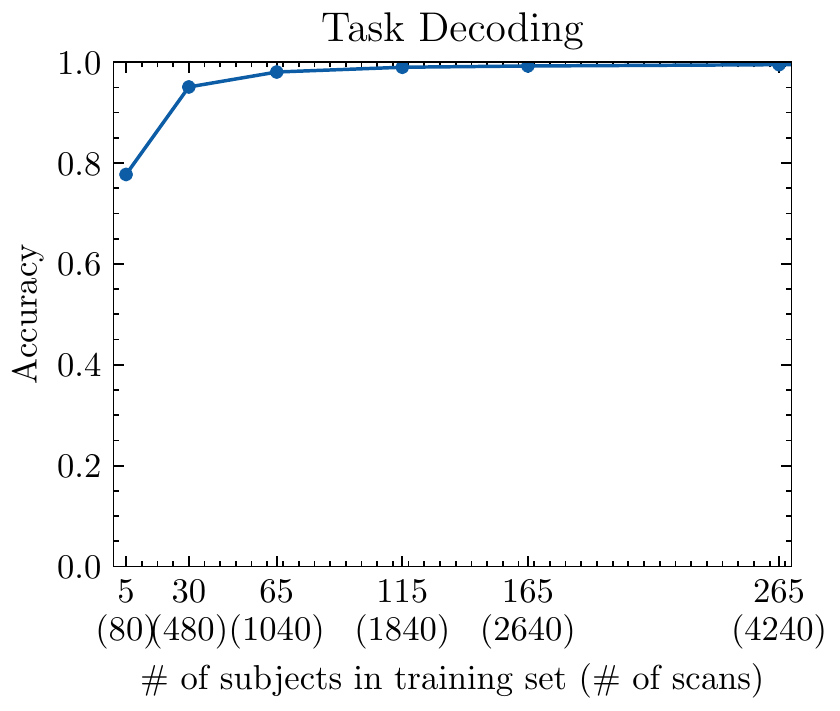}
\caption{Effect of training/validation split on LDA classifier accuracy, for the pipeline with 200 brain regions, global signal regression, no task regression, and no truncation.}\label{training_validation_split}
\end{figure}

The accuracy of the LDA classifier appears to be robust to a reasonably reduced number of scans in the training set (see Figure \ref{training_validation_split}). However, the task decoding classifier maintains nearly 80\% accuracy after being trained on only 5 subjects (80 scans total), whereas for a similar accuracy of around 80\% the fingerprinting classifier requires a training set of a minimum of 2 tasks (3460 scans).

\subsection{Modularity and Centrality}

There has been recent interest in the performance of purely graph-theoretic metrics in functional connectome research. As a point of comparison, we investigated the use of modularity and centrality metrics (instead of the connectome edge weights) as inputs to the best performing classifier architectures across both tasks.

From the 200-region parcellation, a community structure was calculated using Louvain community detection, using asymmetric treatment of negative weights~\cite{rubinov2011weight}. From that community, the participation coefficients were calculated for each brain region, and both positive and negative participation coefficients were concatenated as input ($n=400$ features) to an LDA or NN classifier.

\begin{figure}[H]
\centering
\includegraphics[scale=0.35]{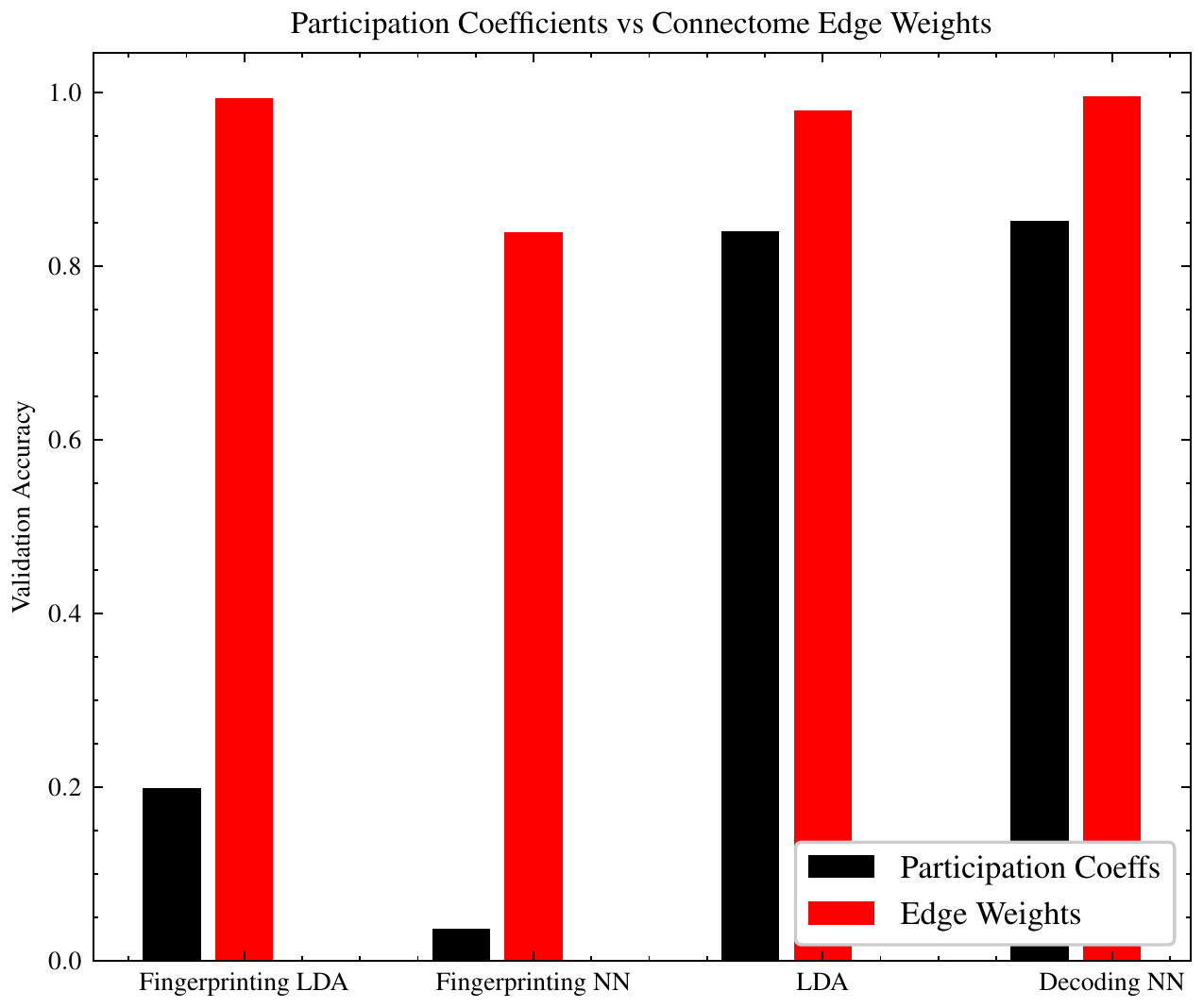}
\caption{Comparison of the classification accuracy achieved by LDA and NN architectures using brain region participation coefficients as input, versus connectome edge weights as input.}
\end{figure}

In both the subject fingerprinting and task classification objective, using the connectome edge weights as input achieves better accuracy than the participation coefficients, though more significantly so in the case of subject fingerprinting.

\subsection{Pre-processing Pipelines}

In this section we present results of experiments comparing the effect of the four pre-processing variables on the classifier accuracies.

\subsubsection{Global Signal Regression}

The number of resting state functional connectivity MRI studies continues to expand at a rapid rate along with the options for data processing. One  such processing option is global signal regression (GSR), that while sometimes controversial, presents several established benefits (see, for example,~\cite{murphy_Towards_a_consensus_GSR} and the references within).

\begin{figure}[H]
\centering
\includegraphics[scale=0.62]{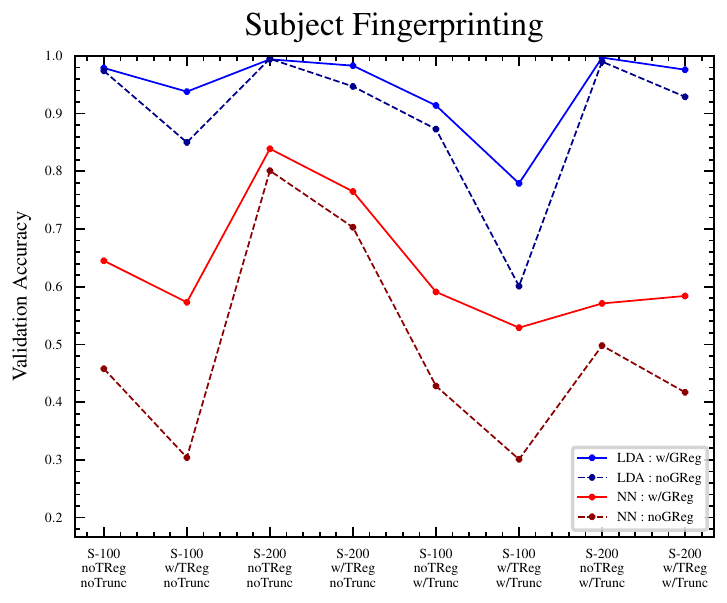}\hspace{1cm}
\includegraphics[scale=0.62]{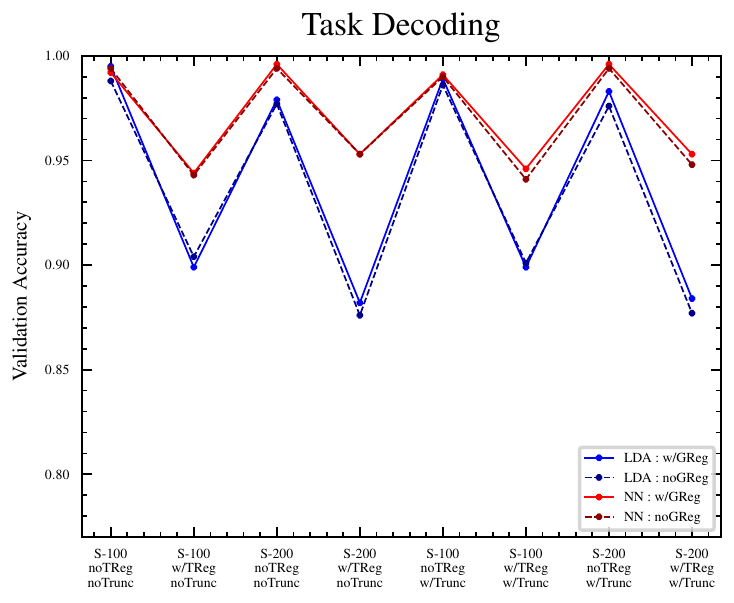}
\caption{With versus without global signal regression.}\label{global_signal}
\end{figure}

For the full set of subjects, global signal regression produced accuracy scores either very similar to pipelines without global signal regression, or else scores that were slightly higher, with one exception: in the case of the NC classifier for the task decoding objective, global signal regression produced scores that were significantly higher overall (see Figure \ref{global_signal}).

\subsubsection{Parcellation Size}

\begin{figure}[H]
\centering
\includegraphics[scale=0.62]{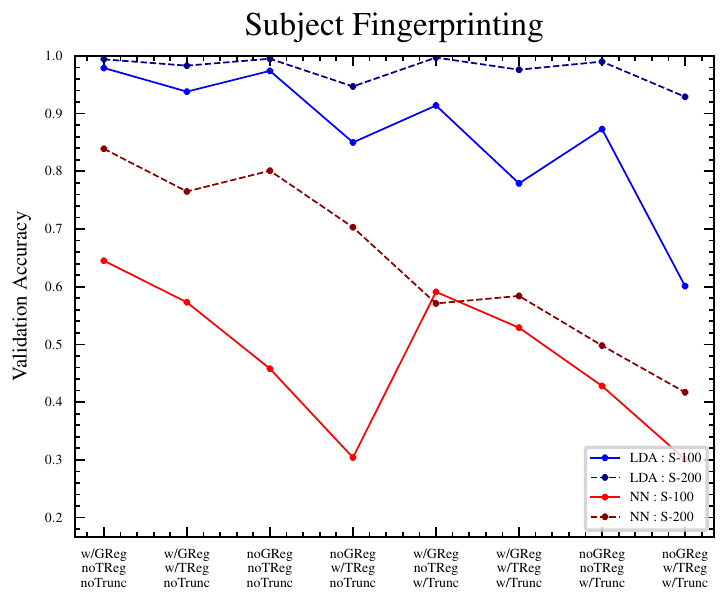}\hspace{1cm}
\includegraphics[scale=0.62]{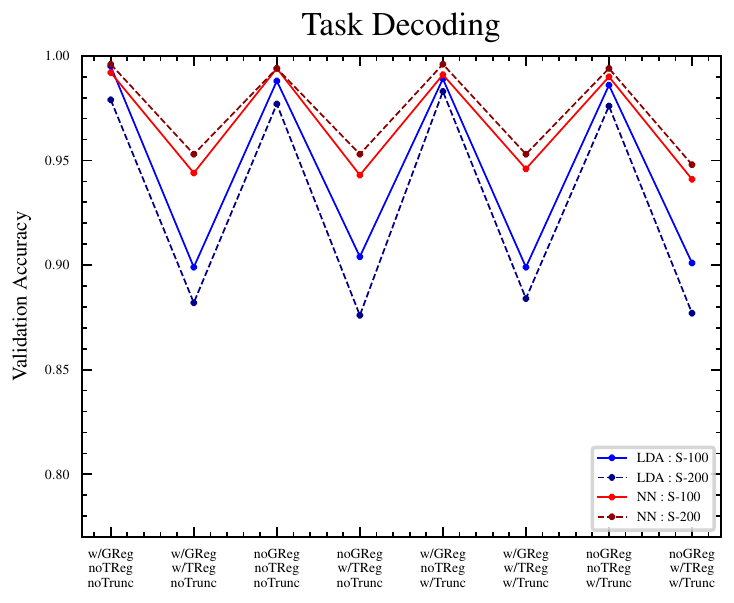}
\caption{100-region versus 200-region parcellation.}\label{parcellation}
\end{figure}

The 100-region and 200-region parcellations resulted in similar accuracy scores, with the 100-region having slightly lower accuracy scores across all pipelines and architectures, with one exception: in the case of the LDA classifier for the task decoding objective, the 100-region parcellation had slightly higher accuracies across all pipelines (see Figure \ref{parcellation}). Aside from that exception, the slight reduction in accuracy is not surprising given that the 200-region parcellation effectively contains the 100-region parcellation as a "subset" of its features. 

\subsubsection{Time-Series Truncation}

Truncation is an important step since it has been established that connection weights stabilize only after ~30m of data (see~\cite[Figure 4]{laumann2015functional}). In general, shorter samples lead to greater sampling error in estimates of connectivity. To ensure that number of samples are comparable we test truncation of the time series to preserve approximately the same number of frames across all samples.

Overall, the truncation of the \emph{rest} scan time series to match the task scans appears to result in slightly poorer accuracy across all pipelines and classifier architectures (see Figure \ref{parcellation}). The deleterious effect is much more pronounced in the 100-region parcellation compared to the 200-region parcellation.

\begin{figure}[H]
\centering
\includegraphics[scale=0.62]{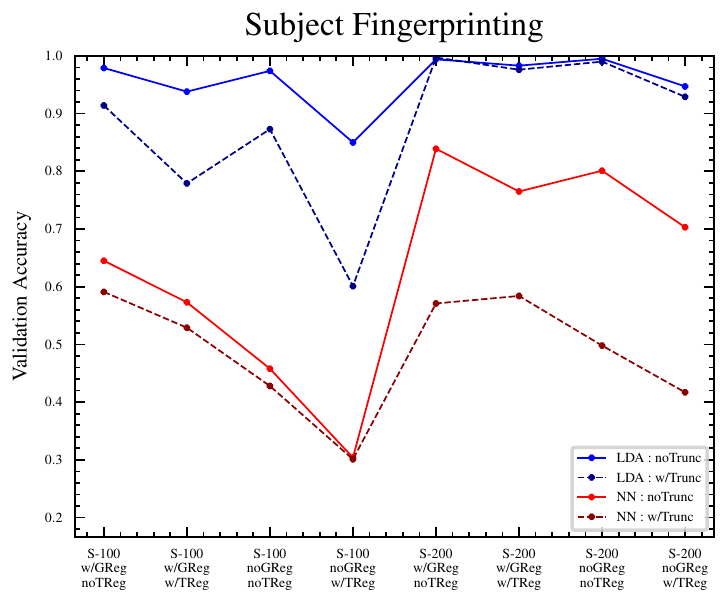}\hspace{1cm}
\includegraphics[scale=0.62]{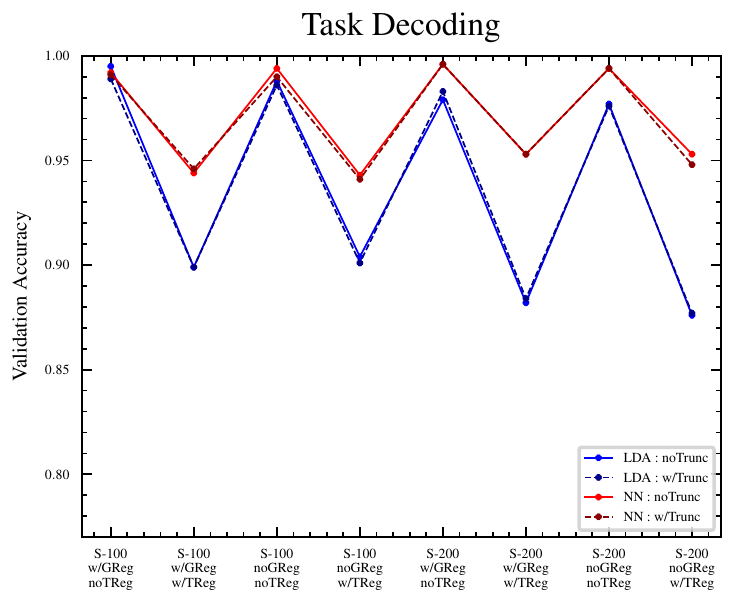}
\caption{With versus without time-series truncation.}\label{truncation}
\end{figure}

\subsubsection{Task Regression}

\begin{figure}[H]
\centering
\includegraphics[scale=0.62]{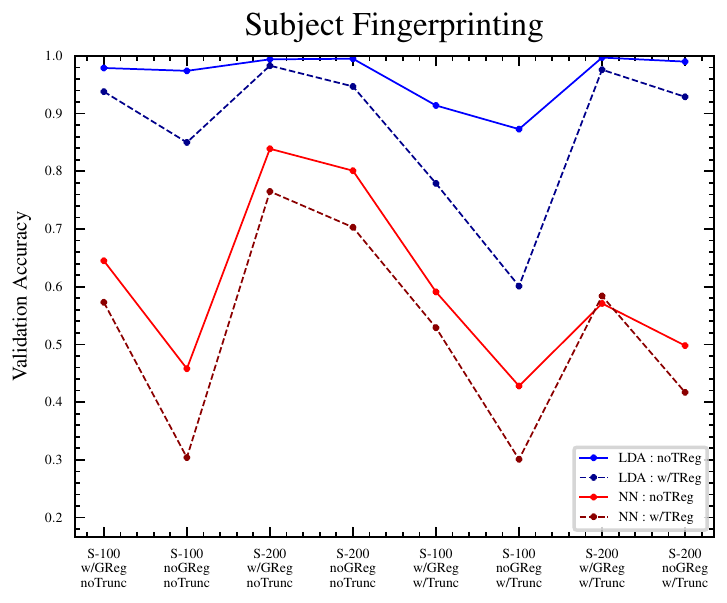}\hspace{1cm}
\includegraphics[scale=0.62]{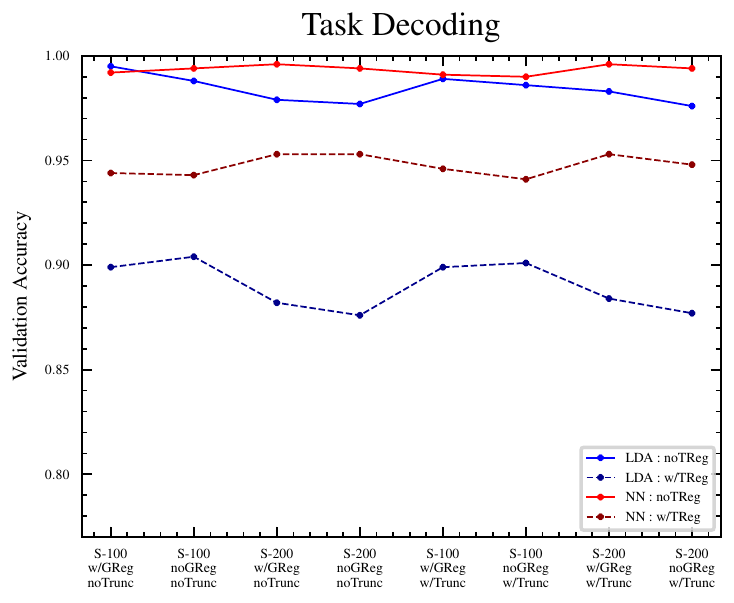}
\caption{With versus without task regression.}\label{task_regression}
\end{figure}

In all tests, pipelines with task regression resulted in poorer classification accuracy than without task regression (see Figure \ref{task_regression}). In the task classification objective this should be somewhat expected - the regression is intended to remove confounds that are associated with each task but that are not true measures of brain region activity. The presence of these task-specific confounds may have made distinguishing between tasks easier, and their removal would be expected to make the task decoding more difficult.

However, we also see a drop in the accuracy scores in the subject fingerprinting objectives with task regression. It is not clear why the removal of task-specific confounds should affect the difficulty of identifying the owner of a scan. It may be that the task regression used is unintentionally removing some important distinguishing information that is not task-specific.

\subsection{Feature Selection}

\begin{figure}[H]
\includegraphics[scale=0.8]{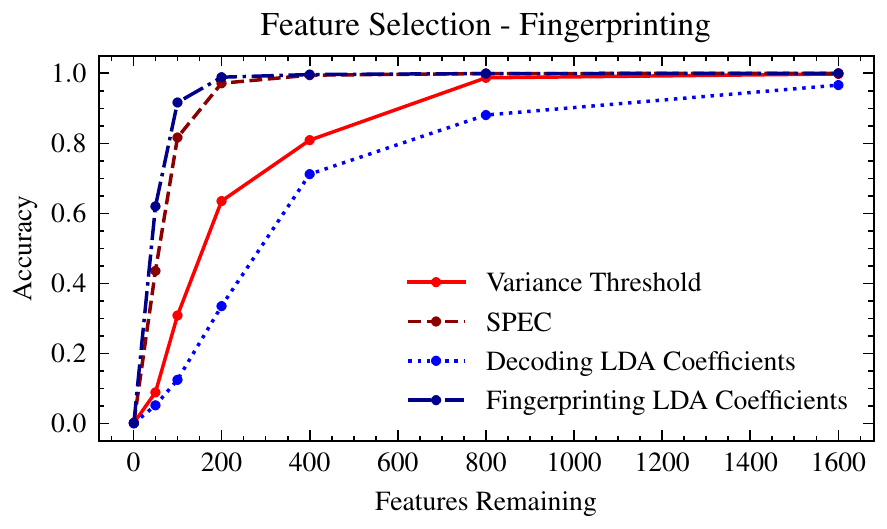}\hspace{1cm}
\includegraphics[scale=0.8]{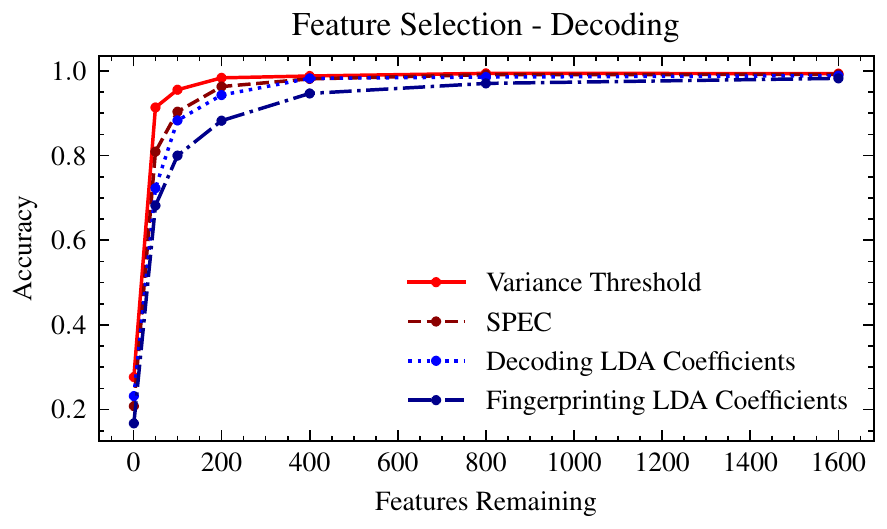}
\caption{Accuracy of LDA classifier (200-region,w/GReg,noTReg) trained on either the fingerprinting or decoding objective, using subsets of the features selected by each of the four feature selection methods.}\label{feature_selection_comp}
\end{figure}

We investigated three methods of feature selection in an attempt to identify edges of the functional connectome that are consistently useful for fingerprinting and decoding objectives:

\begin{itemize}
    \item Variance Threshold: unsupervised method where the variance of a given feature within the entire dataset is used as a measure of its discriminatory power, with the most important features considered to be those that have the highest variance.
    \item SPEC: unsupervised method where graph spectral analysis is used to identify the features that are most important for maintaining the spatial clustering properties of the dataset (see~\cite{Zhao07}).
    \item LDA Coefficients: supervised method where the average of the coefficient of a given feature across all classes for a trained classifier is used as a measure of its discriminatory power, with the most important features considered to be those with the highest average magnitude of their coefficient values.
\end{itemize}

\begin{figure*}
\centering
\includegraphics[scale=0.59]{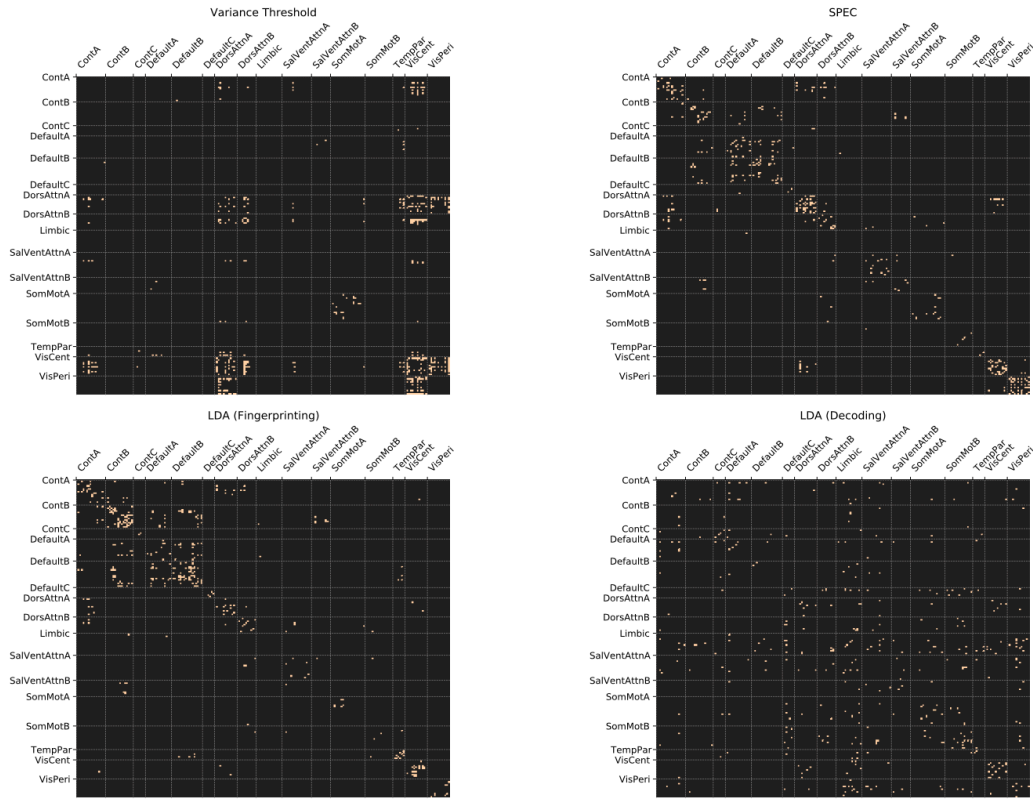}
\caption{Top 200 most important connectome edges selected using each feature selection method, sorted by brain systems.}\label{feature_selection_result}
\end{figure*}

All three feature selection methods were applied to the entire dataset, after which the feature-reduced FCs were tested for accuracy using the normal testing/training split classification objectives using the highest-accuracy pre-processing pipeline (200-region,w/GReg,noTReg).

Results of the feature selection experiments (see Figure \ref{feature_selection_comp}) suggest that high accuracy performance can be maintained across both fingerprinting and decoding objectives using as little as a few hundred features (i.e. connectome edges) from the original \num[group-separator={,}]{19900} (in the case of the 200-region parcellation).

In particular, for any single feature selection method, the feature rankings obtained from SPEC appear to give the best balance of high accuracy across both the fingerprinting and decoding objectives - reaching accuracy in the range of 90\% with only $\sim$200 features.

The brain systems associated with the highest ranked features obtained from SPEC (see Figure \ref{feature_selection_result}) appear to be somewhat of a balance of the highest ranked features from the variance threshold (DorsAttnA, DorsAttnB, VisCent, VisPeri), which has the best overall performance for decoding, and those from the LDA fingerprinting coefficients (ContA, ContB, DefaultA, DefaultB), which has the best overall performance for fingerprinting.

\subsection{High-dimensional Clustering}

\begin{figure}[H]
\centering
\includegraphics[scale=0.33]{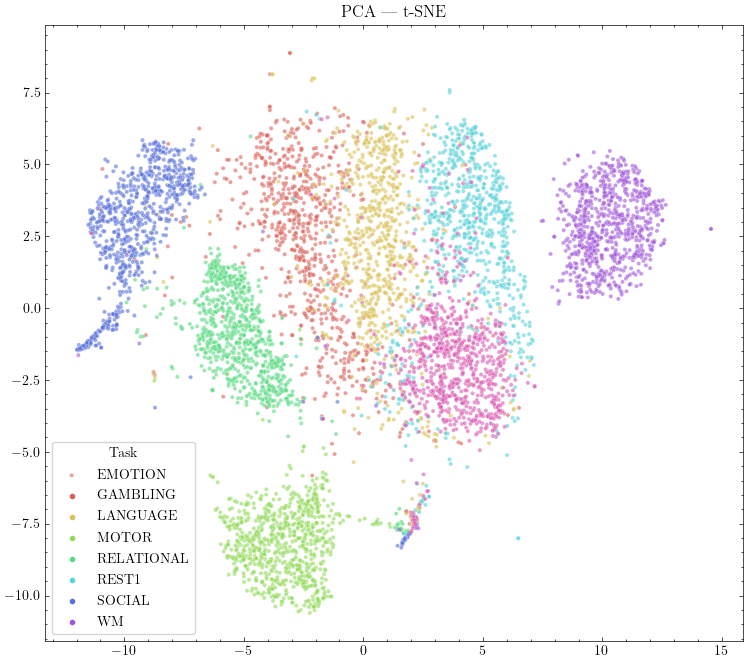}
\caption{Visualization of high-dimensional clustering of cognitive states, following dimensionality reduction using PCA and t-SNE.}\label{clustering}
\end{figure}

As an additional means of evaluating the degree to which classes (individual subjects or tasks) exhibit clustering in high-dimensional space, we performed PCA on the dataset to reduce to 50 dimensions, and then used t-SNE to create a 2-dimensional visualization (see Figure \ref{clustering}).

We also calculated the mean intra-class (within group) and inter-class (between group) distance across all sample pairs, with scans grouped by subject or by task, as a simple measure of clustering (see Figure \ref{class_distance}). Results show that grouping scans by task or by subject both result in a smaller mean intra-class distance than inter-class distance (i.e., on average, scans belonging to an individual, or a particular cognitive state, are closer together than scans between two different subjects or cognitive states).

\begin{figure}[H]
\vspace{-3mm}
\centering
\noindent\begin{tabular}{ ccc  }
\toprule
\thead{Class Grouping} & \thead{Intra-/Inter-class} & \thead{Mean Distance} \\
\cmidrule(lr){1-1}
\cmidrule(lr){2-2}
\cmidrule(lr){3-3}
Subject & Intra-class & 28.250 \\
Subject & Inter-class & 31.389 \\
\cmidrule(lr){1-3}
Task & Intra-class & 30.329 \\
Task & Inter-class & 38.070 \\
\bottomrule
\end{tabular}
\caption{Measure of mean intra-class and inter-class distance for subjects and tasks.}\label{class_distance}
\end{figure}

\subsection{High-Uniformity Subjects}

Additionally, for the subject fingerprinting objective, two smaller subsets of the subjects were tested for robustness against a higher degree of uniformity between subjects:

\begin{itemize}
\item A subset of 100 subjects who had the lowest average motion, as measured by their mean head motion across all scans.
\item A subset of 100 subject (labelled HCP100UR), known to have higher phenotypic uniformity than the dataset as a whole.
\item The HCP100UR subjects were also tested with motion-based frame-censoring to remove high-motion frames, further increasing the uniformity of their scans. For each frame of the scan in which head motion exceeded 0.15 relative RMS, that frame, as well as the two frames before and the two frames after the supra-threshold motion, were removed from the time series. If, for any scan, the number of frames remaining after frame-censoring was less than 50\% of the total, that scan was thrown out. For any subject who did not have all scans available, the subject was not included in the experiment, in order to avoid an unbalanced dataset.
\end{itemize}

Frame-censoring and time-series truncation were not tested together, due to their similarities. The results are summarized in Figure~\ref{HU}.

\begin{figure}[H]
\centering
\footnotesize
\noindent\begin{tabular}{ ccccccc  }
 \toprule
 \multicolumn{4}{c}{Pipeline} & \multicolumn{3}{c}{Accuracy}\\
 \cmidrule(lr){1-4}
 \cmidrule(lr){5-7}
 \thead{Regions} & \thead{Global\\Reg.} & \thead{Task\\Reg.} & \thead{Trunc.} & \thead{Low-\\Movement} & \thead{HCP100UR} & \thead{HCP100UR\\(Frame-Censored)}\\
 \cmidrule(lr){1-4}
 \cmidrule(lr){5-7}
100 & Yes & No   & No   & 0.990 & 0.995 & 0.963\\
100 & Yes & Yes  & No   & 0.955 & 0.940 & 0.900\\
100 & No  & No   & No   & 0.990 & 1.000 & 0.963\\
100 & No  & Yes  & No   & 0.970 & 0.951 & 0.925\\
 \cmidrule(lr){1-4}
 \cmidrule(lr){5-7}
200 & Yes & No   & No   & 1.000 & 0.984 & 0.963\\
200 & Yes & Yes  & No   & 0.985 & 0.962 & 0.850\\
200 & No  & No   & No   & 1.000 & 1.000 & 0.975\\
200 & No  & Yes  & No   & 0.990 & 0.989 & 0.900\\
 \cmidrule(lr){1-4}
 \cmidrule(lr){5-7}
100 & Yes & No   & Yes   & 0.985 & 0.973 & -   \\
100 & Yes & Yes  & Yes   & 0.895 & 0.940 & -   \\
100 & No  & No   & Yes   & 0.985 & 0.967 & -   \\
100 & No  & Yes  & Yes   & 0.955 & 0.951 & -   \\
 \cmidrule(lr){1-4}
 \cmidrule(lr){5-7}
200 & Yes & No   & Yes  & 0.995 & 0.995 & -    \\
200 & Yes & Yes  & Yes  & 0.970 & 0.962 & -    \\
200 & No  & No   & Yes  & 0.990 & 0.995 & -    \\
200 & No  & Yes  & Yes  & 0.995 & 0.989 & -    \\
\bottomrule
\end{tabular}
\caption{Validation accuracy scores of LDA subject fingerprinting classifier on subsets of subjects with higher uniformity.}\label{HU}
\end{figure}










\subsubsection{Low-Movement Subjects.} Limiting the dataset to only the 100 subjects with the lowest average motion resulted in no, or a slight, improvement of the validation accuracy scores across all pipelines. Some of that improvement can likely be attributed to the added ease of classifying between 100 subjects instead of between 865. Nevertheless, these results seem to suggest that average subject movement may not be a major confounding factor.

\subsubsection{HCP100UR Subjects.} Limiting the dataset to only 100 subjects with known high phenotype uniformity resulted in no, or a slight, improvement of the validation accuracy scores across all pipelines. Again, some of that improvement can likely be attributed to an easier classification objective, but as before these results seem to suggest that phenotype uniformity may not be a major confounding factor.

\subsubsection{Movement-Based Frame Censoring.} The removal of high-motion frames from the time series of the HCP100UR subjects resulted in a small overall decrease in accuracy scores. Given that the truncation results already indicate that removal of random frames from the time-series produces poorer accuracy scores, the decrease in accuracy here is not enough to conclude that the high-motion frames in particular are a confounding factor above and beyond the removal of frames in general.

\vspace{-2mm}
\section{Conclusions}

The most notable result is the demonstration that some machine learning techniques and some pre-processing pipelines can achieve high accuracy on both subject fingerprinting and task decoding from FCs, from among a relatively high number of classes in each case: distinguishing between 865 subjects, or between 8 cognitive states, with up to more than 99\% accuracy.

For the best overall performance of a single architecture, the LDA classifier shows a strong ability to discriminate between individual subjects and cognitive states, a consistency of the results across pre-processing pipelines, and provides a useful metric of the means of its classifications via its coefficients.

For subject fingerprinting among a large number of subjects, the LDA classifier gives the highest and most consistent accuracy scores. For task decoding, the LDA, NN, and SVM classifiers perform similarly, though the SVM has a slight advantage across all pipelines.

The pipelines that are most consistently associated with higher classification accuracy scores are: finer parcellations (more brain regions), with global signal regression, without task regression, and without truncation of time-series.

For future applications, when deciding what pre-processing pipeline to use it is important to note that, though a finer parcellation (i.e. more regions) appears to lead to better accuracy, the number of features for a given scan increases with the square of the number of regions. This means that a finer parcellation, though improving accuracy, could potentially make the training of a classifier significantly more computationally expensive. The other aspects of the pipeline investigated here (global regression, task regression, and time-series length) may have their own fixed cost relative to the size of the data set, but should not have an effect on the cost of training a classifier.

A limitation of our work is whether or not it generalizes to other datasets. The high accuracy achieved here is with the benefit of the HCP dataset, which has a high standard of consistency. Future work should investigate whether these results are robust to the inclusion of scans from other datasets - either in the training of the classifiers, or whether the classifiers trained on the HCP dataset maintain their accuracy when applied to scans from other datasets.

\bibliographystyle{plain}
\bibliography{main}

\end{document}